\documentstyle[12pt]{article}
\def\be{\begin{eqnarray}}
\def\ee{\end{eqnarray}}
\def\ba{\begin{array}}
\def\ea{\end{array}}
\def\p{\phi}

\def\a{\alpha}
\def\lam{\lambda}
\def\Lam{\Lambda}
\def\laI{\lambda_{1i}}
\def\laII{\lambda_{2i}}

\def\epII{\epsilon_2}
\def\epI{\epsilon_1}
\def\pa{\partial}

\def\G{{\cal G}}
\def\B{{\cal B}}
\def\A{{\cal A}}
\def\X{{\cal X}}

\def\D{^{(D)}}

\begin{document}
\begin{center}
{\LARGE
{Charged Dual String Vacua from Interacting Rotating \\
\vskip 0.3cm Black Holes Via Discrete and Nonlinear Symmetries}}
\end{center}

\vskip 1.5cm

\begin{center}
{\bf \large {Alfredo Herrera--Aguilar}}
\end{center}

\begin{center}
Instituto de F\'\i sica y Matem\'aticas\\
Universidad Michoacana de San Nicol\'as de Hidalgo\\
Edificio C--3, Ciudad Universitaria, Morelia, Mich., CP 58040, M\'exico\\
e-mail: herrera@zeus.umich.mx
\end{center}

\begin{center}
{\large {and}}
\end{center}

\begin{center}
{\bf \large {Marek Nowakowski}}
\end{center}

\begin{center}
Departamento de F\'\i sica, Universidad de los Andes, \\
Cra. 1 No. 18A--10, Santa Fe de Bogot\'a, Colombia \\
e-mail: mnowakos@uniandes.edu.co
\end{center}

\begin{abstract}
Using the stationary formulation of the toroidally
compactified heterotic string theory in terms of a pair of matrix
Ernst potentials we consider the four--dimensional
truncation of this theory with no $U(1)$ vector fields excited.
Imposing one time--like Killing vector permits us to express the
stationary effective action as a model in which gravity is coupled
to a matrix Ernst potential which, under certain parametrization,
allows us to interpret the matter sector of this theory as a
double Ernst system. We generate a web of string vacua which are
related to each other via a set of discrete symmetries of the
effective action (some of them involve $S$--duality
transformations and possess non--perturbative character). Some
physical implications of these discrete symmetries are analyzed
and we find that, in some particular cases, they relate {\it
rotating} black holes coupled to a dilaton with no Kalb--Ramond
field, {\it static} black holes with non--trivial dilaton and
antisymmetric tensor fields, and rotating and static naked
singularities. Further, by applying a nonlinear symmetry, namely,
the so--called normalized Harrison transformation, on the seed
field configurations corresponding to these neutral backgrounds,
we recover the $U(1)^n$ Abelian vector sector of the
four--dimensional action of the heterotic string, charging in this
way the double Ernst system which corresponds to each one of the
neutral string vacua, i.e., the {\it stationary} and  the {\it
static} black holes and the naked singularities.

\end{abstract}

\newpage
\section{Introduction}
During the last years the concept of duality has played an
important role in the understanding of non--perturbative aspects
of string/M--theory \cite{sen0}. Duality symmetries were
originally discovered for toroidal compactifications of closed
string theories (for a review see \cite{gpr} and references
therein). Subsequently it has been realized that they are a
property of all string vacua for which the metric and the world
sheet have Abelian isometries \cite{buscher}. Afterward, another
symmetry emerged from theories compactified to four and lower
dimensions: the $S$--duality \cite{s}; it involves an exchange of
the electric and magnetic components of the vector fields and
relates the strong and weak coupling regimes of the same/different
string theories. It was precisely the appearance of the $T$-- and
$S$-- dualities that provided the scenario for the correct
understanding of the so--called $U$--duality, which contains the
$T$-- and $S$-- dualities as subgroups \cite{u}.

The existence of these dualities represents another step in the
classification of physically inequivalent string vacua
\cite{witten}. In this context it is interesting to study
solutions of Einstein's equations in the vacuum since they are
also solutions of the leading order string background equations
with constant dilaton and antisymmetric tensor fields \cite{nssw}.
By applying duality to these solutions one generates new
physically different string vacua. This approach has been used in
the framework of the four--dimensional effective Einstein--Maxwell
Dilaton--Axion theory (and some related models; see for example
\cite{emda}--\cite{fvm}) as well as in the context of the world
sheet action for the bosonic string in a background with $N$
commuting isometries \cite{comm}), further generalized to the case
of non--abelian isometries (see \cite{xofq}--\cite{nonab} and
references therein).

In the framework of heterotic string theory toroidally compactified
from $D$ to three dimensions, the use of the symmetry groups of these dualities led
to the formulation of the effective action in terms of a pair of matrix Ernst potentials
(MEP) coupled to gravity \cite{hk3}. Afterward, in
\cite{aha1} it was shown that, under certain parametrization, the matter sector of
the stationary effective action of the truncated $D=4$ heterotic string theory with
one time--like Killing vector and with no vector fields excited (electromagnetic fields
set to zero) can be expressed in terms of a double Ernst system. We briefly review the
MEP formalism and this parametrization in Section 2.

It turns out that, in the language of the Ernst potentials, this
effective action has some discrete symmetries that can be used to
relate a web of physically different string vacua which
occasionally define distinct effective theories in the sense that
they possess different field spectra. Some of these discrete
symmetries involves transformations in which the four--dimensional
dilaton field inverts its sign, corresponding in this way to
$S$--dualities that possess non--perturbative character. Thus, the
string vacua mentioned above are dual to each other since they
have the same mathematical description, but describe distinct
physical configurations. We present this web of dual theories and
field configurations in Section 3 and analyze its physical
meaning. We study the axisymmetric case when the double Ernst
system corresponds to a double Kerr black hole in Section 4. The
field configurations representing interacting black holes have
been recently studied in the framework of both General Relativity
and string theory \cite{2bh}.

By applying a nonlinear symmetry , namely, the normalized Harrison
transformation (NHT) on a neutral field configuration which
corresponds to the double Ernst system, one recovers the $U(1)^n$
Abelian sector that was previously set to zero \cite{aha1}; this
fact corresponds physically to charging the double Ernst system
and, in particular, the double Kerr black hole in the axisymmetric
case. In Section 5 we recall this result in order to further apply
it to the whole web of dual field configurations related via the
discrete symmetries mentioned above and interpret the obtained
charged exact solutions. We summarize our results and discuss
their physical implications in Section 6. Finally, in the
Appendices {\bf A} and {\bf B} we clarify the $S$--duality
character of some of the discrete symmetries that relate the
different string vacua mentioned above.
\section{The effective action and matrix Ernst potentials}
We start with the $D$--dimensional effective action of the
heterotic string at tree level (see \cite{kir}, for instance) \be
S\D\!=\!\int\!d\D\!x\!\mid\!
G\D\!\mid^{\frac{1}{2}}\!e^{-\p\D}\!(R\D\!+\!
\p\D_{;M}\!\p^{(D);M} \!-\!\frac{1}{12}\!H\D_{MNP}H^{(D)MNP}\!-\!
\frac{1}{4}F^{(D)I}_{MN}\!F^{(D)IMN}), \ee where \be
F^{(D)I}_{MN}\!=\!\pa_MA^{(D)I}_N\!-\!\pa _NA^{(D)I}_M, \quad
H\D_{MNP}\!=\!\pa_MB\D_{NP}\!-\!\frac{1}{2}A^{(D)I}_M\,F^{(D)I}_{NP}\!+\!
\mbox{{\rm \, cycl \, perms \,\, of} \,\, M,\,N,\,P.} \nonumber
\ee Here $G\D_{MN}$ is the metric, $B\D_{MN}$ is the
anti--symmetric Kalb-Ramond field, $\p\D$ is the dilaton,
$A^{(D)I}_M$ is a set of $U(1)$ vector fields ($I=1,\,2,\,...,n$)
and $M,N,P=1,2,...,D$. In the consistent critical case $D=10$ and
$n=16$, but we shall leave these parameters arbitrary for the time
being and will consider the $D=4$ theory later on, in Subsec. 2.1.
In \cite{ms}--\cite{sen434} it was shown that after the
compactification of this model on a $D-3=d$--torus, the resulting
stationary theory possesses the $SO(d+1,d+1+n)$ symmetry group
($U$--duality); more precisely, the scheme of the dimensional
reduction contains two steps: one first compactifies the original
theory down to four dimensions on a torus (compact manifold) and
then imposes a time--like Killing vector, i.e., stationarity. It
is worth noticing that for stationary solutions, the equations of
motion of this theory are the same as those of the
$D$--dimensional heterotic string theory for field configurations
independent of $D-3$ dimensions (the time and $D-4$ internal
coordinates), since the character of the analysis is local (see,
for instance, \cite{sen440}).

Thus, this stationary effective field theory describes gravity with the metric tensor
\be
g_{\mu\nu}\!=\!e^{-2\p}\!\left(G\D_{\mu\nu}\!-
\!G\D_{m+3,\mu}G\D_{n+3,\nu}G^{mn}\right),
\ee
coupled to the following set of three--fields:

\noindent a) scalar fields \be G\!\equiv\!G_{mn}\!=
\!G\D_{m+3,n+3},\,\,\, B\!\equiv\!B_{mn}\!= \!B\D_{m+3,n+3},\,\,\,
A\!\equiv\!A^I_m\!= \!A^{(D)I}_{m+3},\,\,\,
\p\!=\!\p\D\!-\!\frac{1}{2}{\rm ln|det}\,G|, \label{escalars} \ee
\noindent b)antisymmetric tensor field \be
B_{\mu\nu}\!=\!B\D_{\mu\nu}\!\!-\!4B_{mn}A^m_{\mu}A^n_{\nu}\!-\!
2\!\left(A^m_{\mu}A^{m+d}_{\nu}\!-\!A^m_{\nu}A^{m+d}_{\mu}\right),
\ee \noindent c)vector fields $A^{(a)}_{\mu}=
\left((A_1)^m_{\mu},(A_2)^{m+d}_{\mu},(A_3)^{2d+I}_{\mu}\right)$
\be (A_1)^m_{\mu}\!=\!\frac{1}{2}G^{mn}G\D_{n+3,\mu},\,
(A_3)^{I+2d}_{\mu}\!=\!-\frac{1}{2}A^{(D)I}_{\mu}\!+\!A^I_nA^n_{\mu},\,
(A_2)^{m+d}_{\mu}\!=\!\frac{1}{2}B\D_{m+3,\mu}\!\!-\!B_{mn}A^n_{\mu}\!+\!
\frac{1}{2}A^I_{m}A^{I+2d}_{\mu} \ee where the subscripts
$m,n=1,2,...,d$; $\mu,\nu=1,2,3$; and $a=1,...,2d+n$. In this
paper we set $B_{\mu\nu}=0$ to remove the effective cosmological
constant from our consideration.

All vector fields in three dimensions can be dualized on--shell
\cite{hk3},\cite{sen434}:
\begin{eqnarray}
\nabla\times\overrightarrow{A_1}&=&\frac{1}{2}e^{2\p}G^{-1}
\left(\nabla u+(B+\frac{1}{2}AA^T)\nabla v+A\nabla s\right),
\nonumber                          \\
\nabla\times\overrightarrow{A_3}&=&\frac{1}{2}e^{2\p}
(\nabla s+A^T\nabla v)+A^T\nabla\times\overrightarrow{A_1},
\label{dual}\\
\nabla\times\overrightarrow{A_2}&=&\frac{1}{2}e^{2\p}G\nabla v-
(B+\frac{1}{2}AA^T)\nabla\times\overrightarrow{A_1}+
A\nabla\times\overrightarrow{A_3}.
\nonumber
\end{eqnarray}
Thus, the effective stationary theory describes gravity
$g_{\mu\nu}$ coupled to the scalars $G$, $B$, $A$, $\p$ and
pseudoscalars $u$, $v$, $s$. These matter fields can be arranged
in the following pair of matrix Ernst potentials: \be \X= \left(
\ba{cc} -e^{-2\p}+v^TXv+v^TAs+\frac{1}{2}s^Ts&v^TX-u^T \cr
Xv+u+As&X \ea \right), \quad \qquad \A=\left(
\ba{c} s^T+v^TA \cr
A \ea \right), \ee where $X=G+B+\frac{1}{2}AA^T$. These matrices
have dimensions $(d+1) \times (d+1)$ and $(d+1) \times n$,
respectively. In terms of the MEP the effective stationary theory
adopts the form \be ^3S\!= \!\int\!d^3x\!\mid
g\mid^{\frac{1}{2}}\!\{\!-\!^3\!R\!+ \!{\rm
Tr}[\frac{1}{4}\left(\nabla \X\!-\!\nabla \A\A^T\right)\!\G^{-1}
\!\left(\nabla \X^T\!-\!\A\nabla \A^T\right)\!\G^{-1}
\!+\!\frac{1}{2}\nabla \A^T\G^{-1}\nabla \A]\}, \label{acXA} \ee
where $\X=\G+\B+\frac{1}{2}\A\A^T$, then \,
$\G=\frac{1}{2}\left(\X+\X^T-\A\A^T\right)$ and \be \G= \left(
\ba{cc} -e^{-2\p}+v^TGv&v^TG \cr Gv&G \ea \right), \quad \B=\left(
\ba{cc} 0&v^TB-u^T \cr Bv+u&B \ea \right). \ee Interestingly there
exist a map between the stationary actions of the heterotic string
and Einstein--Maxwell theories \cite{hk3}: \be
\X\longleftrightarrow -E, \quad \A\longleftrightarrow F, \nonumber
\ee \be {\it matrix\,\,
transposition}\quad\longleftrightarrow\quad {\it complex\,\,
conjugation}, \ee where $E$ and $F$ are the complex Ernst
potentials (gravitational and electromagnetic, respectively) of
the stationary Einstein--Maxwell theory \cite{e}. This map allows
us to extrapolate the results obtained in the EM theory to the
heterotic string one using the MEP formulation.

\subsection{Bosonic string truncation and double Ernst system}
In \cite{aha1} the stationary bosonic string sector of the whole
$D=4$ effective field theory of the heterotic string was
formulated in terms of a double Ernst system coupled to
three--dimensional gravity. This truncated action arises by
dropping the matrix $\A$ in the action (\ref{acXA}), i.e. by
setting to zero all the $U(1)$ Abelian vector fields which
correspond to the winding modes of the reduced theory (we will
recover these Abelian fields later on in the paper by using the
so-called normalized Harrison transformation). Right now we have
\be ^3S\!=\!\int\!d^3x\!\mid
g\mid^{\frac{1}{2}}\!\left\{\!-\!^3\!R\!+ \frac{1}{4}\!{\rm
Tr}\,\left[ \nabla \X\!\G^{-1}\!\nabla
\X^T\!\G^{-1}\right]\right\} =\!\int\!d^3x\!\mid
g\mid^{\frac{1}{2}}\!\left\{\!-\!^3\!R\!+ \frac{1}{4}\!{\rm
Tr}\,\left(J^{\X}J^{\X^{T}}\right)\right\} \label{acX} \ee where
now $\X=\G+\B$,\,\, $\G=\frac{1}{2}\left(\X+\X^T\right)$ and
$J^{\X}=\nabla \X\G^{-1}$. For this theory, since $D=d+3$, then
$d=1$ and $\X$ is a $2\times 2$--matrix which can be fully
parameterized in the following form \be \X= \frac{1}{p_2} \left(
\ba{cc} p_1&p_1q_2-p_2q_1\cr p_1q_2+p_2q_1&p_1(p_2^2+q_2^2)
\ea\right). \label{X} \ee By substituting (\ref{X}) into
(\ref{acX}), the action of the matter fields takes the form
\begin{eqnarray}
^3 S_m = \frac{1}{2}
\int d^3x {\mid g \mid}^{\frac {1}{2}} \left\{
p_1^{-2}\left[(\nabla p_1)^2 + (\nabla q_1)^2\right] +
p_2^{-2}\left[(\nabla p_2)^2 + (\nabla q_2)^2\right]
\right\},
\label{pp}
\end{eqnarray}
which allows us to introduce two independent Ernst--like potentials
\begin{equation}
\epsilon _1 =p_1+iq_1, \qquad \epsilon _2=p_2+iq_2.
\end{equation}
Thus, in terms of these field variables, the action of the field system can
be rewritten in the form
\begin{eqnarray}
^3 S =
\int d^3x {\mid g \mid}^{\frac {1}{2}} \left\{ - ^3 R +
2\left(J^{\epsilon _1}J^{\overline\epsilon _1} +
J^{\epsilon _2}J^{\overline\epsilon _2} \right) \right\},
\label{ace}
\end{eqnarray}
where $J^{\epsilon _1}=\nabla\epsilon _1\,(\epsilon
_1+\overline\epsilon _1)^{-1}$ and $J^{\epsilon _2}=\nabla\epsilon
_2\,(\epsilon _2+\overline\epsilon _2)^{-1}$, which precisely
coincides with the action of a double Ernst system expressed in
the K\"ahler form in the framework of General Relativity. In the
particular axisymmetric case, the double Ernst system describes a
pair of interacting rotating black holes located on the symmetry
axis. However, the action (\ref{ace}) can be obtained for any {\it
stationary} theory of the form (\ref{acX}) with the potential $\X$
parameterized in the form (\ref{X}).

\section{Dual String Vacua Via Discrete Symmetries}
A mathematically equivalent $2 \times 2$--matrix representation of
the three--dimensional effective theory arises from (\ref{pp}) by
making use of the discrete symmetry\footnote {The dual nature of
this discrete symmetry is clarified in the Appendix {\bf A}. In
the next Sec. we shall consider such a duality within the
framework of two interacting rotating black holes of Kerr type.}
$p_1\longleftrightarrow p_2$, \, $q_1 \longleftrightarrow q_2$, a
fact that allows us to define the new matrix potential
\begin{eqnarray}
\X'=
\frac{1}{p_1}
\left (\begin{array}{crc}
p_2& \quad &p_2q_1-p_1q_2\\
p_1q_2+p_2q_1 & \quad &p_2(p_1^2+q_1^2)\\
\end{array} \right),
\label{X'}
\end{eqnarray}
and to write down the action corresponding to these quantities
\be
^3S\!=
\!\int\!d^3x\!\mid g\mid^{\frac{1}{2}}\!\left\{\!-\!^3\!R\!+
\frac{1}{4}\!{\rm Tr}\,\left(J^{\X'}J^{\X'^{T}}\right)\right\}
\!=\!\int\!d^3x\!\mid g\mid^{\frac{1}{2}}\!\left\{\!-\!^3\!R\!+
2\left(J^{\epsilon '_1}J^{\overline\epsilon '_1}+
J^{\epsilon '_2}J^{\overline\epsilon '_2}\right)\right\},
\label{acX'}
\ee
where similarly $J^{\X'}=\nabla\X'\G'^{-1}$, \,
$J^{\epsilon '_1}=\nabla\epsilon '_1\,(\epsilon '_1+\overline\epsilon
'_1)^{-1}$,\,
$J^{\epsilon '_2}=\nabla\epsilon '_2\,(\epsilon '_2+\overline\epsilon
'_2)^{-1}$,\,
$\epsilon '_1=p_2+iq_2$ and $\epsilon '_2=p_1+iq_1$.

In terms of the MEP and complex Ernst potentials, the above--mentioned
discrete transformation reads
\be
\X[\epI,\epII] \longleftrightarrow \X'[\epII,\epI],
\qquad \epsilon_1 \longleftrightarrow \epsilon_1'=\epsilon_2,
\qquad \epsilon_2 \longleftrightarrow \epsilon_2'=\epsilon_1.
\label{map1}
\ee
From the point of view of the three--dimensional quantities, this map
relates the unprimed
\be
G_{tt}
=\frac{Re\epsilon_1}{Re\epsilon_2}\mid\epsilon_2\mid^2, \qquad
B\equiv0, \qquad
e^{-2\p}=-\frac{Re\epsilon_1 Re\epsilon_2}{\mid \epsilon_2\mid ^2}, \qquad
u=Im\epsilon_1, \quad
v=\frac{Im\epsilon_2}{\mid\epsilon_2\mid^2}
\label{unprimed}
\end{eqnarray}
and primed \be
G'_{tt}=\frac{Re\epsilon_2}{Re\epsilon_1}\mid\epsilon_1\mid^2,
\qquad B'\equiv0, \qquad e^{-2\p'}=-\frac{Re\epsilon_1
Re\epsilon_2}{\mid \epsilon_1\mid ^2}, \qquad u'=Im\epsilon_2,
\qquad v'=\frac{Im\epsilon_1}{\mid\epsilon_1\mid^2}
\label{1primed} \ee field configurations parameterized in terms of
the stationary double Ernst system. By looking carefully at this
symmetry one realizes that it mixes the gravitational and matter
degrees of freedom of both theories, since the potentials $p_k$
and $q_k$ ($k=1,2$) enter the matrix $\X$ in a non--symmetric way,
even if they appear in the action in a completely symmetric form.
Namely, under this correspondence, the $G_{t\varphi}$--component
of the four--dimensional metric is directly related to the
$B^{(4)}_{t\varphi}$--component of the Kalb--Ramond field, since
from one side, the pseudoscalar field $u$ ($u'$) is directly
related to the component $G_{t\varphi}$ ($G'_{t\varphi}$) of the
metric (which defines the rotation of the gravitational field)
and, from the other, $v$ ($v'$), defines the component
$B^{(4)}_{t\varphi}$ ($B^{'(4)}_{t\varphi}$) of the antisymmetric
tensor field, (this identification is established through the
dualization relations (\ref{dual})).

The relationship between the $G_{t\varphi}$ and
$B^{(4)}_{t\varphi}$ tensor components also takes place when
applying a Buscher transformation ($T$--duality) on a given
stationary solution of the low--energy string equations
\cite{buscher}. However, in the framework of that discrete
transformation the dilaton field is just shifted and cannot change
its sign. Thus, the strong and weak coupling regimes of the theory
cannot be related each other as in the framework of the symmetry
(\ref{map1}) (see Appendix {\bf A} for details).

A similar effect was found by Bakas in \cite{bakas} as well, where
the implemented $Z_2$ discrete symmetry interchanges the field
content of an $SL(2,{\bf R})/U(1)$ Ernst $\sigma$--model which
describes the gravitational sector of the theory with another
$SL(2,{\bf R})/U(1)$ Ernst $\sigma$--model which parameterizes the
axidilaton one. However, the approach of Bakas is quite different
from the one we present here since the four--dimensional action
and field equations considered in \cite{bakas} were dimensionally
reduced down to two dimensions in the presence of two commuting
space--like Killing symmetries with an ansatz in which $G_{iA}=0$
(i=1,0; A=2,3). Thus, the symmetry group which arises in the
effective two--dimensional theory turns out to be infinite (it
corresponds to the string Geroch group) and the theory itself,
integrable (as it happens within the effective two--dimensional
theory of General Relativity with a similar ansatz).

In the framework of our approach, the discrete symmetry
(\ref{map1}) arises after imposing stationarity (a single
time--like Killing symmetry) and lives in an effective
three--dimensional world, where the vector fields $G_{t\varphi}$
and $B^{(4)}_{t\varphi}$ are non--trivial (they generate the
pseudoscalar fields $u$ and $v$, respectively). Certainly, the
discrete map (\ref{map1}) also relates two non--linear
$\sigma$--models where the gravitational and matter degrees of
freedom are already mixed due to the three--dimensional character
of the effective theory. However, it is remarkable that both
approaches are quite similar since both discrete transformations
relate two non--linear $\sigma$--models, even if they are defined
in different two-- and three--dimensional effective theories. In
this context, it is interesting to consider the further reduction
of the theory down to two dimensions in order to study the
relation of the discrete symmetry (\ref{map1}) to the infinite
dimensional string Geroch group found by Bakas \cite{bakas}.
However, this topic requires a separate investigation.

Finally, it should be noticed that, under the transformation
(\ref{map1}), the dilaton fields $\p_1^{(4)}$ and $\p_2^{(4)}$ are
just interchanged. This means that, for example, {\it rotating}
solutions of the Kaluza--Klein--Dilaton theory are dual to {\it
static} configurations of the bosonic string theory with
nontrivial Kalb--Ramond field of dipole type. This fact will be
illustrated in the next Section.

Another discrete symmetry which is present in the effective three--dimensional
action (\ref{acX}) relates the matrix potential $\X$ to its inverse $\X^{-1}$,
giving rise to the double--primed matrix potential
\begin{eqnarray}
\X''=\X^{-1}=
\frac{1}{p_2(p_1^2+q_1^2)}
\left (\begin{array}{crc}
p_1(p_2^2+q_2^2)& \quad &p_2q_1-p_1q_2\\
-(p_1q_2+p_2q_1)& \quad &p_1\\
\end{array} \right),
\label{X''}
\end{eqnarray}
which defines an action identical to (\ref{acX}) and (\ref{acX'}),
but in terms of the matrix variable $\X''$ (the dual character of
this symmetry is pointed out in Appendix {\bf B}).

In terms of the MEP and the complex Ernst potentials this symmetry reads
\be
\X\longleftrightarrow\X''=\X^{-1},
\qquad \epsilon_1\longleftrightarrow\epI''=\epsilon_1^{-1},
\qquad \epsilon_2\longleftrightarrow\epII''=\epsilon_2^{-1},
\label{map2}
\ee
whereas the three--dimensional double--primed fields written in the language
of the Ernst potentials adopt the form
\be
G''_{tt}=\frac{Re\epsilon_1}{Re\epsilon_2\mid\epsilon_1\mid^2}, \quad
B''\equiv0, \quad
e^{-2\p''}=-\frac{Re\epsilon_1 Re\epsilon_2}{\mid \epsilon_1\mid ^2}, \quad
u''=-\frac{Im\epsilon_1}{\mid\epsilon_1\mid^2},  \quad
v''=-Im\epsilon_2.
\label{2primed}
\ee

Thus, the transformation (\ref{map2}) relates $\epI$ and $\epII$
to their inverse quantities as well. In the particular case when
the imaginary parts of these potentials vanish (static
configurations with no antisymmetric Kalb--Ramond field since $u$,
$u''$, $v$ and $v''$ are set to zero), this map establishes a
correspondence between black holes and naked singularities since
it relates the $G_{tt}$--component to its inverse
$G''_{tt}=G^{-1}_{tt}$. Such a relationship also takes place when
comparing three--dimensional primed and double--primed fields
since the $G'_{tt}$--component also transforms into its inverse
$G''_{tt}=G'^{-1}_{tt}$. This effect is well known in the
framework of General Relativity since it translates horizons into
naked singularities and was pointed out for the first time in the
framework of string theory in \cite{dvv}, when applying
$T$--duality on a given solution (see also
\cite{fvm}--\cite{xofq}). It will be illustrated in the next
Section with an explicit example.

Both symmetries (\ref{map1}) and (\ref{map2}) can be combined in
order to determine another discrete symmetry, namely \be
\X\longleftrightarrow\X'''=\X'^{-1}, \qquad
\epsilon_1\longleftrightarrow\epI'''=\epI'^{-1}=\epII^{-1}, \qquad
\epsilon_2\longleftrightarrow\epII'''=\epII'^{-1}=\epI^{-1},
\label{map3} \ee where the role of the matrix potential is now
played by the matrix
\begin{eqnarray}
\X'''=\X'^{-1}=
\frac{1}{p_1(p_2^2+q_2^2)}
\left (\begin{array}{crc}
p_2(p_1^2+q_1^2)& \quad &p_1q_2-p_2q_1\\
-(p_2q_1+p_1q_2)& \quad &p_2\\
\end{array} \right)
\label{X'''}
\end{eqnarray}
which also defines an action of the form (\ref{acX}) in the
language of $\X'''$. The three--dimensional triple--primed fields
written with the aid of the Ernst potentials are \be
G'''_{tt}=\frac{Re\epsilon_2}{Re\epsilon_1\mid\epsilon_2\mid^2},
\quad B'''\equiv0, \quad e^{-2\p'''}=-\frac{Re\epsilon_1
Re\epsilon_2}{\mid \epsilon_2\mid ^2}, \quad
u'''=-\frac{Im\epsilon_2}{\mid\epsilon_2\mid^2},  \quad
v'''=-Im\epsilon_1. \label{3primed} \ee In particular, when
applied on an original string background, the transformation
(\ref{map3}) combines the effects of the symmetries (\ref{map1})
and (\ref{map2}) since it mixes the gravitational and matter
degrees of freedom of the dual configurations (as before, this
fact can be seen by switching off one of the Ernst potentials, let
us say $\epII=1$) and interchanges the $G_{tt}$--component with
its inverse $G'''_{tt}=G^{-1}_{tt}$, i.e., it relates {\it
rotating} black holes with no Kalb--Ramond field to {\it static}
naked singularities coupled to non--trivial antisymmetric tensor
field. However, this discrete symmetry does not constitute a
$S$--duality transformation since the four--dimensional dilaton
field does not change its sign under it (see Appendix {\bf B} for
details).

We have established in this way a web of discrete symmetries
between physically different string vacua which sometimes
correspond to distinct theories (in the sense that they contain
distinct field spectra) sharing the same mathematical description
in the language of their effective three--dimensional actions.

\section{Double Kerr Solution and Explicit String Vacua}
In this Section we shall give some simple explicit examples of the
dual string vacua related by the discrete symmetries established
above. For concreteness let us consider the axisymmetric double
Kerr black hole system, where the line element in the
Lewis--Papapetrou form reads \be
ds^2=G_{tt}\left(dt+2(A_1)_{\varphi}d\varphi\right)^2+
e^{2\p}\left[e^{2\gamma}\left(d\rho^2+dz^2\right)+\rho^2d\varphi^2\right],
\label{metric} \ee where $\gamma$, $\p$, $G_{mn}$ and
$(A_1)_{\varphi}$ are $\varphi$--independent; moreover, the
function $\gamma$, that accounts for the general relativistic
interaction between the black holes, is $\gamma \equiv \gamma
^{\epsilon _1} + \gamma ^{\epsilon _2}$, whereas the Ernst
potentials are
\begin{eqnarray}
\epsilon_k = 1 - \frac {2m_k}{r_k + ia_k\cos\theta_k},
\end{eqnarray}
where $m_k$ and $a_k$ are constant parameters which define the masses and
rotations of the sources of the Kerr field configurations, respectively. The
Weyl and Boyer--Lindquist coordinates are related through the relations
\begin{eqnarray}
\rho=\sqrt{(r_k - m_k)^2-\sigma _k^2}\sin\theta_k,
\qquad
z = z_k + (r_k - m_k)\cos\theta_k,
\end{eqnarray}
where $z_k$ stands for the positions of the sources, $\sigma _k^2 = m_k^2 - a_k^2$
and, finally,  for the function $\gamma _k$ we have
\begin{equation}
e^{2\gamma _k} = \frac {P_k}{Q_k},
\end{equation}
with
\begin{eqnarray}
P_k &=& \Delta_k - a_k^2 \sin^2 \theta _k, \nonumber \\
Q_k &=& \Delta_k +\sigma _k^2 \sin^2 \theta _k, \nonumber \\
\Delta _k &=& r_k^2 - 2m_k r_k + a_k^2.
\end{eqnarray}
The relationship (\ref{map1}) can be illustrated by switching off
one of the potentials, let us say $\epII =1$; therefore, we get a
four--dimensional field configuration that corresponds to a single
{\it rotating} black hole located at the point $z_1$ with
vanishing dilaton and Kalb--Ramond fields (see Eqs. (2)--(6)),
namely: \be
ds_4^2=G_{tt}\left(dt+\omega_{\varphi}d\varphi\right)^2+
\frac{e^{2\p^{(4)}}}{\mid G_{tt}\mid}
\left[P_1\left(\frac{dr_1^2}{\Delta_1}+d\theta_1^2\right)+
\Delta_1\sin^2\theta_1d\varphi^2\right], \nonumber \ee \be
G_{tt}=-\frac{(r_1^2-2m_1r_1+\a_1^2\cos^2\theta_1)}
{(r_1^2+\a_1^2\cos^2\theta_1)}, \qquad \omega_{\varphi}\equiv
2(\vec{A_1})_{\varphi} =\frac{2m_1\a_1r_1\sin^2\theta_1}
{(r_1^2-2m_1r_1+\a_1^2\cos^2\theta_1)}, \qquad B\equiv 0,
\nonumber \ee \be e^{2\gamma}
=\frac{(r_1^2-2m_1r_1+\a_1^2\cos^2\theta_1)}
{(r_1^2-2m_1r_1+\a_1^2\cos^2\theta_1+m_1^2\sin^2\theta_1)}, \qquad
\p^{(4)}=0, \qquad B^{(4)}_{t\varphi}\equiv 2(\vec{A_2})_{\varphi}
=0. \label{sb} \ee Under the discrete symmetry (\ref{map1}) this
solution is mapped into a {\it static} black hole configuration
endowed with non--trivial dilaton and Kalb--Ramond field of dipole
type: \be ds_4^{'2}=G'_{tt}dt^2+\frac{e^{2\p^{'(4)}}}{\mid
G'_{tt}\mid}
\left[P_1\left(\frac{dr_1^2}{\Delta_1}+d\theta_1^2\right)+
\Delta_1\sin^2\theta_1d\varphi^2\right], \qquad
\omega'_{\varphi}\equiv 2(A_1)_{\varphi}=0, \qquad B'\equiv 0,
\nonumber \ee \be
G'_{tt}=-\frac{(r_1^2-4m_1r_1+4m_1^2+\a_1^2\cos^2\theta_1)}
{(r_1^2-2m_1r_1+\a_1^2\cos^2\theta_1)}, \qquad e^{2\gamma}
=\frac{(r_1^2-2m_1r_1+\a_1^2\cos^2\theta_1)}
{(r_1^2-2m_1r_1+\a_1^2\cos^2\theta_1+m_1^2\sin^2\theta_1)},
\nonumber \ee \be B^{'(4)}_{t\varphi}\equiv 2(A_2)_{\varphi}
=\frac{2m_1\a_1\sin^2\theta_1(r_1-2m_1)}
{(r_1^2-2m_1r_1+\a_1^2\cos^2\theta_1)}, \quad \p^{'(4)}={\rm
ln}\frac{(r_1^2-4m_1r_1+4m_1^2+\a_1^2\cos^2\theta_1)}
{(r_1^2-2m_1r_1+\a_1^2\cos^2\theta_1)}. \label{sb1} \ee Thus, the
known duality between the rotational sector of the metric
$\omega_{\varphi}$ and the $B^{(4)}_{t\varphi}$--component of the
Kalb--Ramond field (see \cite{buscher} and \cite{bakas}) also
takes place under the discrete symmetry (\ref{map1}). A similar
effect takes place when we switch off the potential
$\epsilon_1=-1$. However, it is worth noticing that if we do not
switch off any potential $\epsilon_k$, the sources of the
four--dimensional dilaton fields just exchange their positions
$z_1\longleftrightarrow z_2$, in other words,
$\p^{(4)}(z_1)\longleftrightarrow \p^{(4)}(z_2)$.

The discrete symmetry (\ref{map2}) also acts in a non--trivial way
on a given string background. This fact can be illustrated by
considering, for example, static configurations with no
antisymmetric Kalb--Ramond field (the pseudoscalar fields $u$,
$u''$, $v$ and $v''$ are set to zero, therefore, the imaginary
parts of the Ernst potentials vanish). Thus, the metric of an
original four--dimensional Schwarzschild solution in the Einstein
frame\footnote{The relationship between the metrics in the
Einstein and string frames in four dimensions reads
$ds_E^2=e^{-\phi^{(4)}}ds_{st}^2$.} with the source located at the
position $z_1$ and coupled to the dilaton field
$e^{\phi^{(4)}}=Re\epII=1-2m_2/r_2$ with its source located at
$z_2$ reads \be
ds_E^2=-\left(\frac{r_1-2m_E}{r_1}\right)dt^2+\left(\frac{r_1}{r_1-2m_E}\right)
\left[\left(\frac{r_2^2-2m_2r_2}{r_2^2-2m_2r_2+m_2^2\sin^2\theta_2}\right)\times
\right. \nonumber\ee \be \left.
\left(dr_1^2+\left(r_1^2-2m_Er_1\right)d\theta_1^2\right)
+\left(r_1^2-2m_Er_1\right)\sin^2\theta_1 d\psi^2\right]; \ee its
under (\ref{map2}) this solution defines a geometry represented by
the following metric \be
ds_E^{''2}=-\left(\frac{r_1}{r_1-2m_E}\right)dt^2+\left(\frac{r_1-2m_E}{r_1}\right)
\left[\left(\frac{r_2^2-2m_2r_2}{r_2^2-2m_2r_2+m_2^2\sin^2\theta_2}\right)\times
\right. \nonumber\ee \be \left.
\left(dr_1^2+\left(r_1^2-2m_Er_1\right)d\theta_1^2\right)
+\left(r_1^2-2m_Er_1\right)\sin^2\theta_1 d\psi^2\right],
\label{sb2} \ee with the dilaton field inverted
$\phi^{''(4)}=1/Re\epII=(1-2m_2/r_2)^{-1}$. The
$G_{tt}$--component of the metric also transforms into its
inverse. As pointed out above, this relationship is well known in
the framework of both General Relativity and string theory and
connects black hole geometries with naked singularities. For
example, in the context of the worldsheet action for the bosonic
string in a background with non--abelian isometries, it was shown
that such a relationship maps the horizon into a naked singularity
\cite{xofq} (see as well \cite{dvv}--\cite{fvm}, for instance).
Thus, under the discrete duality symmetry (\ref{map2}), the black
hole geometry is translated into a naked singularity, whereas the
four dimensional dilaton field inverts its sign. A similar effect
can be established by comparing the three--dimensional primed and
double--primed fields. It is interesting to study the origin of
these naked singularities in general and their relation to
$S$--duality transformations. However, this topic deserves a
separate investigation.

As it was pointed out above, the discrete map (\ref{map3})
combines the physical effects of the previous symmetries. In order
to illustrate this result we shall switch off the Ernst potential
$\epII=1$. Thus, under this symmetry, the string background
(\ref{sb}) representing a rotating black hole configuration with
vanishing dilaton and Kalb--Ramond fields transforms into the
following field configuration \be
ds_4^{'''2}=G'''_{tt}dt^2+\frac{e^{2\p^{'''(4)}}}{\mid
G'''_{tt}\mid}
\left[P_1\left(\frac{dr_1^2}{\Delta_1}+d\theta_1^2\right)+
\Delta_1\sin^2\theta_1d\varphi^2\right], \qquad
\omega'''_{\varphi}\equiv 2(A_1)_{\varphi}=0, \qquad B'''\equiv 0,
\nonumber \ee \be
G'''_{tt}=-\frac{(r_1^2+\a_1^2\cos^2\theta_1)}{(r_1^2-2m_1r_1+\a_1^2\cos^2\theta_1)},
\qquad e^{2\gamma} =\frac{(r_1^2-2m_1r_1+\a_1^2\cos^2\theta_1)}
{(r_1^2-2m_1r_1+\a_1^2\cos^2\theta_1+m_1^2\sin^2\theta_1)},
\nonumber \ee \be B^{'''(4)}_{t\varphi}\equiv 2(A_2)_{\varphi}
=-\frac{2m_1r_1\a_1\sin^2\theta_1}
{(r_1^2-2m_1r_1+\a_1^2\cos^2\theta_1)}, \quad \p^{'''(4)}={\rm
ln}\frac{(r_1^2+\a_1^2\cos^2\theta_1)}
{(r_1^2-2m_1r_1+\a_1^2\cos^2\theta_1)}, \label{sb3} \ee which
describes a static metric coupled to non--trivial dilaton and
antisymmetric tensor field of dipole type. By looking at the
behaviour of the $G_{tt}$--component of the metric tensor under
(\ref{map3}), one realizes that it again gets inverted. Hence, the
transformed geometry constitutes a naked singularity. Therefore,
the discrete transformation (\ref{map3}) establishes a
relationship between four--dimensional {\it rotating} black holes
with no matter fields and {\it static} naked singularities coupled
to non--trivial dilaton and Kalb--Ramond fields. However, since
the dilaton field does change its sign under the discrete symmetry
(\ref{map3}), the transformed string background does not involve
or contain a $S$--duality transformation.

Similar relationships can be established between the four families
of string vacua connected via the discrete symmetries
(\ref{map1}), (\ref{map2}) and (\ref{map3}). We emphasize that
each symmetry acts in a non--trivial way on the starting
four--dimensional field configurations.

\section{Charged Field Configurations }
In the language of the MEP the stationary action (\ref{acXA})
possesses a set of symmetries which has been classified according
to their charging properties in \cite{hk5}. Among them, only the
nonlinear Ehlers and Harrison transformations act in a
non--trivial way on the spacetime \cite{har}. For instance, the
so--called normalized Harrison transformation allows us to
construct charged string vacua from neutral ones preserving the
asymptotic values of the seed fields. Namely, the matrix
transformation \be
&&\A\rightarrow\left(1+\frac{1}{2}\Sigma\lambda\lambda^T\right)
\left(1-\A_0\lambda^T+\frac{1}{2}\X_0\lambda\lambda^T\right)^{-1}
\left(A_0-\X_0\lambda\right)+\Sigma\lambda,
\\
&&\X\rightarrow\left(1+\frac{1}{2}\Sigma\lambda\lambda^T\right)
\left(1-\A_0\lambda^T+\frac{1}{2}\X_0\lambda\lambda^T\right)^{-1}
\left[\X_0+\left(\A_0-\frac{1}{2}\X_0\lambda\right)\lambda^T\Sigma\right]
+\frac{1}{2}\Sigma\lambda\lambda^T\Sigma, \nonumber \ee where
$\Sigma=diag(-1,-1;1,1,...,1)$ and $\lambda$ is an arbitrary
constant $(d+1)\times n$--matrix parameter, generates charged
string solutions (with non--zero electromagnetic potential $\A$)
from neutral ones if we start from the seed potentials $\X_0\ne 0$
and $\A_0=0$. Thus, this solution generation procedure allows us
to generate the $U(1)^n$ electromagnetic spectrum of the effective
heterotic string theory starting with just the bosonic string
spectrum. In other words, if we apply the NHT on uncharged seed
solutions (like (\ref{unprimed}) (\ref{1primed}), (\ref{2primed})
or (\ref{3primed}), for instance), we recover the $U(1)^n$ vector
field sector of the heterotic string theory, charging, by this
means, the double Ernst system and, in particular, interacting
Kerr black hole configurations. It is interesting to see what kind
of transformation undergo the unprimed and primed seed solutions
after making use of the NHT.

The seed MEP that correspond to the neutral stationary double
Ernst system are \be \X_0= \left( \ba{ccc}
\frac{p_1}{p_2}&\quad&\frac{p_1q_2-q_1p_2}{p_2}\cr
\quad&\quad&\quad\cr \frac{p_1q_2+q_1p_2}{p_2}&\quad&
\frac{p_1}{p_2}(p_2^2+q_2^2) \ea \right), \qquad \qquad \A_0=0.
\ee For the simplest case the charge matrix $\lam$ that
parameterizes the normalized Harrison transformation has the form
\be \lam= \left( \ba{cccc} \lam_{11}&\lam_{12}&...&\lam_{1n}\cr
\lam_{21}&\lam_{22}&...&\lam_{2n}\cr \ea \right), \ee where $n\ge
2$ for consistency; these parameters $\laI$ and $\laII$ ($i=1,
2,...,n$) can be interpreted as the electromagnetic charges of the
generated field configuration. When $n=6$ the generated field
spectrum corresponds to the bosonic sector of ${\cal N}=4$, $D=4$
supergravity; here we shall leave it arbitrary for the sake of
generality. After applying the normalized Harrison transformation
on a generic double Ernst seed solution with this matrix $\lam$,
the transformed MEP read \be
\X_{11}\!=\!\frac{1}{\Xi}\left[\left(4+\Lam^2 |\epII |^2\right)
Re\epI+2\left(\laI^2+\laII^2 |\epI |^2\right)Re\epII - 4\laI\laII
Re\epII Im\epI\right], \ee \be
\X_{12}\!=\!\frac{1}{\Xi}\left\{\Gamma_{-}\left( Re\epI
Im\epII-Re\epII Im\epI\right)+2\laI\laII\left[ (1-|\epII
|^2)Re\epI +(1-|\epI |^2)Re\epII\right]\right\}, \ee \be
\X_{21}\!=\!\frac{1}{\Xi}\left\{\Gamma_{+} \left(Re\epI
Im\epII+Re\epII Im\epI\right)+2\laI\laII\left[ (1-|\epI
|^2)Re\epII -(1-|\epII |^2)Re\epI\right]\right\}, \ee \be
\X_{22}\!=\!\frac{1}{\Xi}\left[ \left(\Lam^2 +4|\epII |^2\right)
Re\epI+2\left(\laII^2+\laI^2 |\epI |^2\right)Re\epII + 4\laI\laII
Re\epII Im\epI\right], \ee \be \A_{1j}\!=\!\frac{-2}{\Xi}\left\{
\left[\left(2+\laII^2 |\epII |^2\right)Re\epI + \left(2+\laII^2
|\epI |^2\right)Re\epII + \laI\laII\left(Re\epI Im\epII-Re\epII
Im\epI\right)\right]\lam_{1j}+\right. \nonumber \ee \be
\left.\left[\left(2-\laI^2\right)\left(Re\epI Im\epII-Re\epII
Im\epI\right)- \laI\laII\left(|\epII |^2Re\epI +|\epI
|^2Re\epII\right)\right]\lam_{2j} \right\}, \ee \be
\A_{2j}\!=\!\frac{-2}{\Xi}\left\{\left[\left(2-\laII^2\right)
\left(Re\epI Im\epII+Re\epII Im\epI\right)-\laI\laII \left(Re\epI
+|\epI |^2 Re\epII\right)\right]\lam_{1j}\right.+ \nonumber \ee
\be \left.\left[\left(\laI^2+2|\epII |^2\right)Re\epI +
\left(2+\laI^2 |\epI |^2\right)Re\epII + \laI\laII\left(Re\epI
Im\epII+Re\epII Im\epI\right)\right]\lam_{2j}\right\}, \ee \be
\Xi=2\left(\laI^2+\laII^2 |\epII |^2\right)Re\epI + \left(4+\Lam^2
|\epI |^2\right)Re\epII + 4\laI\laII Re\epI Im\epII, \ee where
$\Gamma_{-}=4-2\laI^2+2\laII^2-\Lam^2$,
$\Gamma_{+}=4+2\laI^2-2\laII^2-\Lam^2$ and
$\Lam^2=\laI^2\lambda_{2j}^2-\left(\laI\laII\right)^2$.

Now we can consider the explicit seed solutions corresponding to the field
configurations (\ref{unprimed}), (\ref{1primed}), (\ref{2primed}) and
(\ref{3primed}), substitute the expressions of $\epI$ and $\epII$ in the
transformed MEP formulae, compute the original fields with the aid of
(2)--(6), and give an interpretation of the generated families of
solutions.

After applying the NHT on the seed solution (\ref{unprimed}) we
get the following unprimed three--dimensional field
configurations: \be G\equiv G_{tt}=\X_{22}-\frac{1}{2}\A_{2j}^2,
\qquad B\equiv0, \qquad A=\A_{2j}, \qquad
v=\frac{1}{2G}\left(\X_{12}+\X_{21}-\A_{1j}\A_{2j}\right),
\nonumber \ee \be u=v\X_{22}-\X_{12},  \qquad
s^T=\A_{1j}-v\A_{2j}, \qquad
e^{-2\p}=\frac{1}{2}\left[v(\X_{12}+\X_{21})+\A_{ij}s\right]-\X_{11},
\ee where the appearance of the electromagnetic potential is
obvious. Similar relations hold for the primed, double-- and
triple--primed field systems that arise under the interchanges
(\ref{map1}), (\ref{map2}) and (\ref{map3}), respectively.

With the aid of the dualization relations (\ref{dual}) we get explicit
expressions for the non--trivial components of the vector fields
\be
\omega_{\varphi}\equiv 2(A_1)_{\varphi}
=\left.\left[-4\chi_{11}+2\lam_{1i}^2\chi_{21}+\left(
\lam_{1i}^2\lam_{2j}^2-(\lam_{1i}\lam_{2i})^2\right)\chi_{12}-
2\lam_{2i}^2\chi_{22}+4\lam_{1i}\lam_{2i}\chi_{23}\right]\right/DQ,
\nonumber
\ee
\be
B^{(4)}_{t\varphi}\equiv 2(A_2)_{\varphi}
=\left.\left[-2\lam_{2i}^2\chi_{11}
+\left(\lam_{1i}^2\lam_{2j}^2-(\lam_{1i}\lam_{2i})^2\right)\chi_{21}
+2\lam_{1i}^2\chi_{12}
-4\chi_{22}-4\lam_{1i}\lam_{2i}\chi_{13}\right]\right/DQ,
\nonumber
\ee
\be
2(A_3)^{i'}_{\varphi}
=2(DQ)^{-2}\left\{\lam_{1i}\lam_{2i}\left[DQ\left(\chi_{12}+\chi_{21}\right)-
\left(4-2\lam_{2i}^2\right)\chi_{22}+\left(\lam_{1i}^2\lam_{2j}^2-
2\lam_{1i}^2-(\lam_{1i}\lam_{2i})^2\right)\chi_{21}-\right.\right.
\nonumber
\ee
\be
\left.\left.4\,\lam_{1i}\,\lam_{2i}\,\chi_{13}\right]+
2\left[\left(1-\lam_{1i}^2\right)\left(2-\lam_{2j}^2\right)-
\left(1-\lam_{2i}^2\right)(\lam_{1j}\,\lam_{2j})^2\right]
\left(\chi_{13}-\chi_{23}\right)\right\}\lam_{1i'}+
\nonumber
\ee
\be
2(DQ)^{-2}\left\{2DQ\left(\chi_{11}+\chi_{22}\right)-
DQ\lam_{1i}^2\left(\chi_{12}+2\chi_{21}\right)+
2DQ\lam_{1i}\lam_{2i}\left(\chi_{13}-\chi_{23}\right)-\right.
\nonumber
\ee
\be
\left.
2\lam_{1i}\lam_{2i}\left[\lam_{1j}\lam_{2j}\left(\chi_{21}+\chi_{22}\right)+
\left(4-\lam_{1j}^2-\lam_{2j}^2\right)\chi_{13}+
\left(\lam_{2j}^2-\lam_{1j}^2\right)\chi_{23}\right]\right\}\lam_{2i'},
\ee
where $i'=1,2,...n$ and the functions $\chi_{kl}$ ($l=1,2,3;$) and $DQ$ read
\be
\chi_{k1}=
\frac{\a_km_kr_k\sin^2\theta_k}
{(r_k^2-2m_kr_k+\a_k^2\cos^2\theta_k)},
\qquad
\chi_{k2}=
\frac{\a_km_k(r_k-2m_k)\sin^2\theta_k}
{(r_k^2-2m_kr_k+\a_k^2\cos^2\theta_k)},
\nonumber
\ee
\be
\chi_{k3}=
\frac{m_k(r_k^2-2m_kr_k+\a_k^2)\cos\theta_k}
{(r_k^2-2m_kr_k+\a_k^2\cos^2\theta_k)},
\qquad
DQ=4-2\lam_{1i}^2-2\lam_{2i}^2+
\lam_{1i}^2\lam_{2j}^2-(\lam_{1i}\lam_{2i})^2.
\ee
It is worth noticing that since the $\chi_{k3}$ functions do not
vanish at spatial infinity (they involve the so--called NUT
parameters of the gravitational field), in order to get an
asymptotically flat gravitational
field configuration, i.e., to deal with charged black holes, we should
impose the orthogonality condition on the pair of charge vectors
$\lam_{1i}$ and $\lam_{2i}$:
\be
\lam_{1i}\lam_{2i}=0.
\label{orto}
\ee
However, even with this restriction on the $(A_3)^{i'}_{\varphi}$ fields
the Dirac string singularity is still present; in order to remove it we
could, in principle, normalize the charge vector $\lam_{1i}$:
\be
\lam_{1i}^2=1,
\ee
but we shall keep it unnormalized for the sake of generality (in fact, this
singularities could correspond to the Dirac strings of monopole--type
solutions).

Thus, after imposing the conditions (\ref{orto}), the transformed
metric preserves its form (\ref{metric}) with the following field
configurations in the string frame
\be
G_{tt}=-\frac{(2-\lam_{2i}^2)^2P_1P_2
\left[4R_2\tilde r_1+\lam_{1j}^4R_1\tilde r_2-4\lam_{1j}^2\left(P_1P_2-
4m_1m_2\a_1\a_2\cos\theta_1\cos\theta_2\right)\right]}
{\left[-2P_1\left(\lam_{1j}^2\tilde r_2+\lam_{2i}^2R_2\right)+
P_2\left(4\tilde r_1+\lam_{1j}^2\lam_{2i}^2R_1\right)\right]^2}
\nonumber
\ee
\be
e^{\p^{(4)}}
=\frac{\left(2-\lam_{2i}^2\right)\left[4R_2\tilde r_1
+\lam_{1j}^4R_1\tilde r_2-4\lam_{1j}^2\left(P_1P_2-
4m_1m_2\a_1\a_2\cos\theta_1\cos\theta_2\right)\right]}
{\left(2-\lam_{1j}^2\right)\left[-2P_1\left(\lam_{1j}^2
\tilde r_2+\lam_{2i}^2R_2\right)+P_2\left(4\tilde r_1+
\lam_{1j}^2\lam_{2i}^2R_1\right)\right]}
\nonumber
\ee
\be
\omega_{\varphi}\equiv 2(A_1)_{\varphi}
=\left.\left(-4\chi_{11}+2\lam_{1j}^2\chi_{21}+
\lam_{1j}^2\lam_{2i}^2\chi_{12}-2\lam_{2i}^2\chi_{22}\right)\right/
\left[\left(2-\lam_{1j}^2\right)\left(2-\lam_{2i}^2\right)\right],
\nonumber
\ee
\be
B^{(4)}_{t\varphi}\equiv 2(A_2)_{\varphi}
=\left.\left(-2\lam_{2i}^2\chi_{11}+\lam_{1j}^2\lam_{2i}^2\chi_{21}
+2\lam_{1j}^2\chi_{12}-4\chi_{22}\right)\right/
\left[\left(2-\lam_{1j}^2\right)\left(2-\lam_{2i}^2\right)\right],
\nonumber
\ee
\be
2(A_3)^{i'}_{\varphi}
=\left.4\left(1-\lam_{1j}^2\right)\left(\chi_{13}-\chi_{23}\right)
\lam_{1i'}\right/\left(2-\lam_{1j}^2\right)^2+
\nonumber
\ee
\be
\left.\left[4\left(\chi_{11}+\chi_{22}\right)-2\lam_{1j}^2
\left(\chi_{12}+2\chi_{21}\right)\right]\lam_{2i'}\right/
\left[\left(2-\lam_{1j}^2\right)\left(2-\lam_{2i}^2\right)\right],
\ee
where $R_k=(r_k-2m_k)^2+\a_k^2\cos^2\theta_k$,
$\tilde r_k=r_k^2+\a_k^2\cos^2\theta_k$ and $e^{2\gamma}$ remains
the same.

\section{Conclusion and Discussion}
We have presented a web of dual four-dimensional bosonic string
vacua that have the same mathematical description in terms of a
matrix Ernst potential. These backgrounds are related by discrete
symmetries and all of them describe distinct physical
configurations. In particular, it turns out that the discrete
symmetry (\ref{map1}) mixes the gravitational and matter degrees
of freedom of the dual backgrounds, i.e., the $G_{t\varphi}$
component of the gravitational metric is interchanged with the
$B^{(4)}_{t\varphi}$ component of the Kalb--Ramond antisymmetric
tensor field, establishing in this way a duality between {\it
rotating} solutions of the Kaluza--Klein--Dilaton theory and {\it
static} configurations of the bosonic string theory with
nontrivial antisymmetric tensor field of dipole type.

Another discrete symmetry (\ref{map2}) establishes a
correspondence between black holes and naked singularities since
it relates the $G_{tt}$--component to its inverse
$G''_{tt}=G^{-1}_{tt}$ (at least for static field configurations).
This interesting non--trivial effect on the singularities of the
dual theories is accompanied by the inversion if the sign of the
four--dimensional dilaton field. Thus, in this particular case,
the discrete symmetry (\ref{map2}) corresponds to a $S$--duality
transformation and possesses non--perturbative character. We
obtain the same physical result by comparing the
three--dimensional primed and double--primed fields, namely, the
$G'_{tt}$--component is also related to its inverse
$G''_{tt}=G'^{-1}_{tt}$ under this symmetry.

A third discrete symmetry arises when one combines the
transformations (\ref{map1}) and (\ref{map2}). Thus, as one could
expect, when applied on a generic string background, the
transformation (\ref{map3}) combines the effects of the symmetries
(\ref{map1}) and (\ref{map2}) and mixes the gravitational and
matter degrees of freedom of the dual configurations and
interchanges the $G_{tt}$--component with its inverse
$G'''_{tt}=G^{-1}_{tt}$, i.e., relates pure {\it rotating} black
hole configurations to {\it static} naked singularities coupled to
non--trivial antisymmetric Kalb--Ramond and dilaton fields.
However, this discrete symmetry does not involve any
transformation which could correspond to $S$--duality since the
four--dimensional dilaton field does not change its sign under it.

These are just some simple examples of the dual field
configurations and theories connected via the discrete symmetries
(\ref{map1}), (\ref{map2}) or (\ref{map3}). Some of the physical
effects that take place under these transformations are of
interest since they have non--perturbative character. It should be
stressed that each symmetry acts in a non--trivial way on the
starting four--dimensional string background.

Finally, by making use of the nonlinear charging symmetry called
normalized Harrison transformation we endow all the generated
string vacua with electromagnetic charges. We clarify the
conditions under which the constructed charged gravitational
configurations are asymptotically flat in order to interpret the
obtained field configurations from (\ref{unprimed}) as charged
black holes.

In the framework of these results, it is of interest as well to
know whether the discrete symmetries presented in this work can
map singular regions into regular regions of one string black hole
geometry (as it happens in \cite{g}) or of different black hole
geometries \cite{gq}, or relate charged black strings to boosted
(uncharged) black strings as in \cite{hhs} after applying the NHT.

In this context, it is interesting to consider the further
reduction of the theory down to two dimensions in order to study
the relation of the discrete symmetries (\ref{map1}) and
(\ref{map2}) to the infinite dimensional string Geroch group found
by Bakas \cite{bakas}.

It would be interesting as well to extend this kind of discrete
dualities to spaces compactified in Calabi--Yau manifold since
they have no continuous isometries (the O(d+1,d+1) symmetry group
in this case) but they are known to have duality--like symmetries
\cite{cogp}. Another interesting issue concerns the possibility of
getting an Ernst system corresponding to $n>2$ alined rotating
sources (black holes, in particular) by means of a suitable
parametrization of the $(d+1)\times (d+1)$--matrix potential $\X$.

\section*{Appendix A}
Let us analyze the dual nature of the discrete symmetry
(\ref{map1}). As pointed out in the Introduction, the
$U$--duality group of the low--energy string theory contains the
$S$-- and $T$--duality groups as subgroups (see \cite{u} and
\cite{sen434}, for instance). The fact that the symmetry
(\ref{map1}) relates the gravitational and matter degrees of
freedom under the transformation found by Buscher
\cite{buscher} indicates that the discrete map (\ref{map1})
contains the $T$--duality symmetry. For this reason, we shall
show below that the symmetry (\ref{map1}) also contains the $S$--duality
transformation since it interchanges the sign of the
four--dimensional dilaton field, and, thus, relates the strong and
weak coupling regimes of the effective theory under consideration.

From relations (\ref{unprimed}) it follows that the Ernst
potentials read \be
\epsilon_1=-\frac{\sqrt{-G_{tt}}}{e^{\phi}}+iu, \qquad \qquad
\epsilon_2=\frac{\sqrt{-G_{tt}}e^{\phi}-iG_{tt}e^{2\phi}v}{1-G_{tt}e^{2\phi}v^2}.
\label{a1} \ee Under the discrete symmetry (\ref{map1}) the
expressions for $\epI$ and $\epII$ interchange and we get \be
\epsilon_1=\frac{\sqrt{-G'_{tt}}e^{\phi'}-iG'_{tt}e^{2\phi'}v'}
{1-G'_{tt}e^{2\phi'}v^{'2}}, \qquad
\epsilon_2=-\frac{\sqrt{-G'_{tt}}}{e^{\phi'}}+iu'. \label{a2} \ee
Thus, the primed three--dimensional fields are related to the
unprimed ones through the following equations \be
G'_{tt}=-\frac{u^2e^{2\phi}-G_{tt}}{1-G_{tt}e^{2\phi}v^2}, \qquad
\qquad\qquad \qquad \qquad
u'=\frac{-G_{tt}e^{2\phi}v}{1-G_{tt}e^{2\phi}v^2}, \nonumber\ee
\be e^{2\phi'}=-\frac{\left(u^2e^{2\phi}-G_{tt}\right)
\left(1-G_{tt}e^{2\phi}v^2\right)}{G_{tt}e^{2\phi}}, \qquad \qquad
v'=\frac{ue^{2\phi}}{u^2e^{2\phi}-G_{tt}}. \label{a3} \ee By
looking at the behaviour of the four--dimensional dilaton field
(see eq. (\ref{escalars})) under such a symmetry we observe that
it changes its sign: \be
e^{\phi^{(4)'}}=e^{\phi'}\sqrt{-G'_{tt}}=\frac{u^2e^{2\phi}-G_{tt}}
{e^{\phi}\sqrt{-G_{tt}}}=\frac{u^2e^{2\phi}-G_{tt}}{e^{\phi^{(4)}}}.
\label{a4} \ee Thus, one can conclude that the discrete symmetry
(\ref{map1}) involves transformations in which the dilaton changes
its sign and, thus, contains the $S$--duality symmetry.

\section*{Appendix B}
As we did in the Appendix {\bf A}, here we shall study the effect
of the discrete symmetry (\ref{map2}) on the four--dimensional
dilaton and show that under certain conditions it also inverts the
sign of the dilaton field. In Appendix {\bf A} we pointed out
that from relations (\ref{unprimed}) it follows that the Ernst
potentials $\epsilon_k$ adopt the form (\ref{a1}) in terms of the
three--dimensional fields. Under the discrete symmetry
(\ref{map2}) the Ernst potentials transform into their inverse
and, thus, we can express them in terms of the double--primed
fields \be
\epsilon_1=-\frac{\sqrt{-G''_{tt}}e^{\phi''}+iu''e^{2\phi''}}
{u''^2e^{2\phi''}-G''_{tt}}, \qquad \qquad
\epsilon_2=\frac{1}{\sqrt{-G''_{tt}}e^{\phi''}}-iv''. \label{b1}
\ee From (\ref{a1}) and (\ref{b1}) we can extract the
relationships that connect the double--primed and unprimed
three--dimensional field variables \be
G''_{tt}=-\frac{e^{-2\phi}-G_{tt}v^2}{u^2-G_{tt}e^{-2\phi}},
\qquad \qquad\qquad \qquad \qquad
u''=\frac{-u}{u^2-G_{tt}e^{-2\phi}}, \nonumber\ee \be
e^{2\phi''}=-\frac{\left(u^2-G_{tt}e^{-2\phi}\right)
\left(e^{-2\phi}-G_{tt}v^2\right)}{G_{tt}e^{-2\phi}}, \qquad
\qquad v''=\frac{G_{tt}v}{e^{-2\phi}-G_{tt}v^2}. \label{b2} \ee By
computing the transformed four--dimensional dilaton field we get
\be
e^{\phi^{(4)''}}=e^{\phi''}\sqrt{-G''_{tt}}=\frac{e^{-2\phi}-G_{tt}v^2}
{e^{-\phi}\sqrt{-G_{tt}}}=e^{-\phi^{(4)}}+v^2e^{\phi^{(4)}}.
\label{b3} \ee This expression indicates us that when considering
field configurations in which $v=0$ we deal with a strong--weak
coupling duality transformation. Thus, the discrete symmetry
(\ref{map2}) contains the $S$--duality.

It should be pointed out as well that the discrete symmetry
(\ref{map3}) does not correspond to a $S$--duality transformation
since the implementation of the discrete symmetry (\ref{map2})
reverts the effect generated by the (\ref{map1}), i.e.,
(\ref{map2}) reverts the inversion of the sign that the
four--dimensional dilaton field received under (\ref{map1}) . Thus, as
a result, the dilaton field does not change its sign under the
discrete transformation (\ref{map3}). This fact can be seen as
follows.

The expressions for the Ernst potentials in terms of the
triple--primed fields are \be
\epsilon_1=\frac{-1}{\sqrt{-G'''_{tt}}e^{\phi'''}}-iv''',
\qquad\qquad
\epsilon_2=\frac{\sqrt{-G'''_{tt}}e^{\phi'''}-iu'''e^{2\phi'''}}
{u'''^2e^{2\phi'''}-G'''_{tt}}. \label{c1} \ee Correspondingly,
the expressions of the triple--primed field variables in terms of
the unprimed ones read \be G'''_{tt}=1/G_{tt}, \quad \qquad
u'''=-v, \qquad\quad e^{\phi'''}=e^{\phi}, \quad \qquad v'''=-u.
\label{c2} \ee As before, we compute the expression for the
transformed four--dimensional dilaton field \be e^{\phi^{(4)'''}}=
\frac{e^{\phi}}{\sqrt{-G_{tt}}}. \label{c3} \ee and observe that
it does not correspond to the original four--dimensional dilaton
field with the inverted sign. Thus, we can conclude that the
discrete symmetry (\ref{map3}) does not involve a $S$--duality
transformation.

\section*{Acknowledgments}
One of the authors (AHA) acknowledges useful discussions with H.H.
Garc\'\i a--Compe\'an, is really grateful to S. Kousidou for
encouraging him during the performance of this research and thanks
IMATE--UNAM and CINVESTAV for library facilities provided when
this work was in progress. The AHA's research was supported by
grants CONACYT-J34245-E and CIC-UMSNH-4.18.


\end{document}